\documentclass[11pt,tightenlines,eqsecnum,floats,aps,amsmath,amssymb,nofootinbib,prd,shownopacs,floatfix]{revtex4}
\usepackage{graphicx}
\usepackage{epstopdf}
\usepackage{latexsym}
\usepackage{amssymb}
\usepackage{amsmath}
\usepackage{color}
\usepackage{mathrsfs}
\usepackage{xparse}
\usepackage{float}
\usepackage{mathtools}
\usepackage{multirow}
\usepackage{physics}
\usepackage[hidelinks]{hyperref}
\hypersetup{
  colorlinks = true, %Colours links instead of ugly boxes
  urlcolor   = cyan, %Colour for external hyperlinks
  linkcolor  = darkgray, %Colour of internal links
  citecolor  = blue %Colour of citations
}
\usepackage{cleveref}
\usepackage[center]{subfigure}

\usepackage{xcolor}
\usepackage[normalem]{ulem}
\usepackage{comment}

\begin{document}
\renewcommand\arraystretch{2}
 \newcommand{\bq}{\begin{equation}}
 \newcommand{\eq}{\end{equation}}
 \newcommand{\bqn}{\begin{eqnarray}}
 \newcommand{\eqn}{\end{eqnarray}}
 \newcommand{\nb}{\nonumber}
 \newcommand{\cb}{\color{blue}}
    \newcommand{\cc}{\color{cyan}}
     \newcommand{\lb}{\label}
        \newcommand{\cm}{\color{magenta}}
\newcommand{\rc}{\rho^{\scriptscriptstyle{\mathrm{I}}}_c}
\newcommand{\rd}{\rho^{\scriptscriptstyle{\mathrm{II}}}_c}
\NewDocumentCommand{\evalat}{sO{\big}mm}{%
  \IfBooleanTF{#1}
   {\mleft. #3 \mright|_{#4}}
   {#3#2|_{#4}}%
}

\newcommand{\PRL}{Phys. Rev. Lett.}
\newcommand{\PL}{Phys. Lett.}
\newcommand{\PR}{Phys. Rev.}
\newcommand{\CQG}{Class. Quantum Grav.}
\newcommand{\parallelsum}{\mathbin{\!/\mkern-5mu/\!}}

\title{Spherically symmetric loop quantum gravity: Schwarzschild spacetimes with a cosmological constant}
%\title{Transient emergent-universe-like phase from the polymerized scalar field and its cosmological impacts}
\author{Esteban Mato$^{1}$}
\email{emato@fing.edu.uy}
\author{Javier Olmedo$^{2}$}
\email{javolmedo@ugr.es}
\author{Sahil Saini $^{3}$}
\email{sahilsaiini@gjust.org}
\affiliation{$^{1}$ Instituto de F\'isica, Facultad de Ingenier\'ia (Universidad de la Rep\'ublica), Julio Herrera y Reissig 565, 11300 Montevideo, Uruguay,\\
$^{2}$ Departamento de F\'isica Te\'orica y del Cosmos, Universidad de Granada, Granada-18071, Spain.\\
$^{3}$ Department of Physics, Guru Jambheshwar University of Science \& Technology, Hisar, Haryana 125001, India\\}

\begin{abstract}
We provide a quantization of the Schwarzschild spacetime in the presence of a cosmological constant, based on midisuperspace methods developed in the spherically symmetric sector of loop quantum gravity, using in particular the 'improved dynamics' scheme. We include both the de Sitter and anti-de Sitter cases. We find that the quantization puts a Planckian positive upper limit on the possible values of the cosmological constant similar to the bounds obtained earlier from studies of homogeneous spacetimes. This means that, for negative cosmological constant, no negative bound is found. Moreover, using semiclassical physical states, we obtain the effective metric and demonstrate the causal structure for various cases. Quantum gravity modifications ensure that the singularity is replaced by a transition surface in all the cases, where the curvature invariants approach mass-independent Planckian bounds. Analysis of the effective stress-energy tensor shows that the null energy condition is strongly violated in the vicinity of the transition surface. Moreover, it shows a weaker asymptotic fall off for a nonvanishing cosmological constant, which could have interesting phenomenological implications.

\end{abstract}

\maketitle

\section{Introduction}
In recent years, while we are still far from directly observing quantum gravity effects, black holes have become a test bed for various theories of quantum gravity. A successful and theoretically satisfactory resolution of the black hole singularity is the first step towards building confidence in quantum gravity models. In this direction, considerable progress has been made in loop quantum gravity (LQG). It has been found through numerous studies, approaching the problem from different directions, that a loop quantization resolves the singularity replacing it by a regular transition surface beyond which the spacetime can be extended. Most of the studies are based on the minisuperspace approach used in loop quantum cosmology, where symmetry-reduction leaves only a finite number of degrees of freedom to be loop quantized. For recent developments using this approach, see \cite{AOSreview,HusainEwing2022,Singh2023}.

In case of spherically symmetric spacetimes, an independent program of loop quantization based on the midisuperspace approximation has been developed. The spherically symmetric framework was initially proposed in \cite{Bengtsson,TK93,BK2000,B2004,BS2006} where symmetry reduction was carried out and kinematical framework was laid down. However, the resulting model turns out to be a nonlinear constrained field theory, with a non-Abelian constraint algebra showing structure functions, as in the full theory. In order to overcome this challenge, a redefinition of the Hamiltonian constraint was carried out to Abelianize the algebra of the Hamiltonian constraint with itself. This redefinition in turn allowed the Dirac quantization of the model in the case of the Schwarzschild black holes \cite{GP2013,CGP2007,GOP2014}. The framework has been further extended \cite{GOP2020,GOP2021} to implement the improved dynamics scheme developed by Chiou et. al. \cite{Chiou2012}, which was adapted to this family of midispuerspace models but within the context of polarized Gowdy cosmologies on the three torus \cite{dBOP2017}. In recent years, this approach has been applied to obtain a quantization of the charged black holes \cite{GMP2015,BMO2024}. 

In this manuscript, we extend the program of spherically symmetric loop quantization to Schwarzschild spacetimes with a cosmological constant $\Lambda$, where $\Lambda$ is kept general and allowed to be both positive or negative. Apart from being of interest in their own right, these models provide an opportunity to study quantum black holes in non asymptotically flat spacetimes. We first Abelianize the classical Hamiltonian constraint by suitable coordinate transformations, and then loop quantize the resulting theory along the lines of \cite{CGP2007,GP2013,GOP2014}. Following \cite{dBOP2017,BMO2024,GOP2020}, we adopt the improved dynamics scheme implementing two separate improved dynamics conditions as in Ref. \cite{BMO2024}, which allows us to determine all quantization parameters. 

We find that the quantum theory imposes a mass and $\Lambda$-dependent finite lower bound on how small the smallest area spheres can be. Furthermore, the leading order term in this lower bound on the radius of smallest area spheres agrees with that obtained for the charged black hole \cite{BMO2024} as well as the uncharged Schwarzschild black hole studied in \cite{GOP2020,GOP2021} in the improved dynamics scheme. In contrast to the charged black hole, where the first order corrections due to charge were negative, in the case of a cosmological constant we found that they were positive. Moreover, we also found that this quantum theory puts a Planckian upper bound on how large a positive cosmological constant can be. This agrees with upper bounds found in homogeneous spacetimes with a non-vanishing positive $\Lambda$ \cite{posLambda-LQC,posLambda-LQC2}. It ensures that the quantum geometry effects limit the curvature in the asymptotic limit, which is proportional to $\Lambda$, to Planckian values at most. Besides, and in agreement with the analysis of homogeneous spacetimes with negative $\Lambda$ \cite{AdS_LQC}, we found no upper bound on the magnitude of negative $\Lambda$. We show all this explicitly in the section on effective geometry. 

In particular, using semiclassical physical states, we obtain the effective metric containing the lowest order quantum corrections. We find that the effective metric is regular at the center and the singularity is replaced by a transition surface which links a trapped region with an anti-trapped region. We analyze the causal structure of the resulting spacetimes - all of which are asymptotically de Sitter or anti-de Sitter. Let us recall that, in the classical theory, and depending on the value of the cosmological constant, there are four distinct spacetimes - (i) $0<9G^2 M^2\Lambda<1$ corresponds to the conventional Schwarzschild-de Sitter spacetime, (ii) $9G^2 M^2\Lambda>1$ corresponds to ultramassive spacetimes \cite{senovilla-23a,senovilla-23b} with a naked singularity at the center, (iii) $9G^2 M^2\Lambda=1$ is the special case where the black hole horizon overlaps with the cosmological horizon, and (iv) $\Lambda<0$ corresponds to the anti-de Sitter case. The spacetime represented by the effective metric in all cases is a regular extension of the corresponding classical spacetimes beyond the central singularity. The analysis of curvature invariants reveals that quantum effects only become significant when the curvature approaches Planckian values. In particular, we show that for macroscopic black holes, the curvature invariants approach a mass-independent Planckian bound at the transition surface. The analysis of the effective stress-energy tensor also shows a strong violation of the null energy condition (NEC) in the vicinity of the transition surface, indicating the resolution of the singularity.

The manuscript is organized as follows. In section II, we provide the classical theory of the symmetry-reduced Schwarzschild spacetime with a cosmological constant and obtain an Abelianized Hamiltonian constraint. Section III contains the kinematical details of loop quantization using the improved dynamics scheme. In section IV on physical states and observables we give the conditions that need to be satisfied for a consistent quantization - which yields a Planckian upper bound on positive $\Lambda$. In section V, we obtain the effective metric and derive its physical implications, including the analysis of causal structure and the behavior of curvature invariants. We summarize our findings in section VI. The details of quantum dynamics of the model are contained in an appendix at the end.

We set the Immirzi parameter $\gamma = 1$ as well as $G=1=\hbar$.

\section{Schwarchild spacetime with a cosmological constant: the classical theory}

The geometrical sector of the theory is described in terms of Ashtekar variables. In spherically symmetric spacetimes, the connection has two gauge invariant components, $(K_\varphi, K_x)$, in the radial and transverse directions, and similarly, the densitized triad has also two gauge invariant components $(E^\varphi,E^x)$ \cite{B2004}. They form two canonical pairs of field variables and fullfil the Poisson algebra
\begin{eqnarray}
\{K_x(x), E^x(x')\}=G\delta (x-x'),\\
\{K_\varphi(x), E^\varphi(x')\}=G\delta (x-x').
\end{eqnarray}
There is no matter content but we will include a cosmological constant $\Lambda$. Therefore, our spacetime will not be asymptotically flat, as we will see. 

The total reduced Hamiltonian is a combination of constraints 
\begin{equation}
    H_{T}=G^{-1}\int dx (NH_{\rm gr}+N^x H_x),
\end{equation}
where the diffeomorphism and scalar constraints are given by,
\begin{eqnarray}
    H_x &=& E^{\varphi}K'_{\varphi} - (E^x)'K_x, \\ \nonumber
    H_{\rm gr} \\ 
    &=& \left(\frac{((E^x)')^2}{8\sqrt{|E^x|}E^{\varphi}} - \frac{E^\varphi}{2\sqrt{|E^x|}} - 2K_\varphi \sqrt{|E^x|}K_x - \frac{E^{\varphi} K^2_\varphi}{2\sqrt{|E^x|}} - \frac{\sqrt{|E^x|}(E^x)'(E^{\varphi})'}{2(E^\varphi)^2}+ \frac{\sqrt{|E^x|}(E^x)''}{2E^\varphi}  \right. \nonumber \\ 
    &&+ \left. \sqrt{|E^x|}E^\varphi \frac{\Lambda}{2} \right).
\end{eqnarray}
The corresponding spacetime metric can be written as
\begin{equation}
\label{metric}
    ds^2= -(N^2 - N_x N^x) dt^2 + 2N_x dt dx + \frac{(E^\varphi)^2}{|E^x|}dx^2 +|E^x| d\Omega^2,
\end{equation}
where $d\Omega^2 = d\theta^2 + \sin^2\theta d\varphi^2$ being the metric of the unit sphere and $N_x=g_{xx} N^x = \frac{(E^\varphi)^2}{|E^x|}N^x$. 

The constraint algebra 
\begin{align}
     &\{H_x(N_x),H_x(\tilde N_x)\}=H_x(N_x\tilde N_x'-N_x'\tilde N_x),\, \\
     &\{H_{\rm gr}(N),H_x(N_x)\}=H_{\rm gr}(N_x N'),\\
     &\{H_{\rm gr}(N),H_{\rm gr}(\tilde N)\}=H_x\left(\frac{|E^x|}{(E^\varphi)^2}\left[N\tilde N'-N' \tilde N\right]\right),
     \end{align}
as usual, involves structure functions. To Abelianize the algebra of the Hamiltonian constraint with itself, we follow the ideas of \cite{CGP2007,GP2013,GOP2014} and introduce new lapse and shift functions as follows
\begin{equation}
    N^x= \bar N^x - 2N\frac{K_\varphi \sqrt{|E^x|}}{(E^x)'}, \quad N=\bar N \frac{(E^x)'}{E^\varphi},
\end{equation}
which changes the scalar constraint while leaving the diffeomorphism constraint unchanged
\begin{eqnarray}
    \bar H(\bar N)&=&-\int dx \bar N \left[ \left(\sqrt{|E^x|} \left(1 + K^2_\varphi -\frac{[(E^x)']^2}{4(E^\varphi)^2} \right) \right)' - \sqrt{|E^x|} (E^x)' \frac{\Lambda}{2} \right],\\
    H_x(\bar N^x)&=& \int dx \bar N^x [-(E^x)'K_x+E^\varphi K'_\varphi ].
\end{eqnarray}
Rewriting $\sqrt{|E^x|} (E^x)' \Lambda/2$ as $[|E^x|^{3/2}\Lambda/3]' $ and then integrating by parts we obtain
\begin{equation}
    H_{\rm ab}(\tilde N)=-\int dx \Tilde{N} \left[-\sqrt{|E^x|}\left(1+K^2_\varphi - \frac{[(E^x)']^2}{4(E^\varphi)^2}\right)+2GM + |E^x|^{3/2}\frac{\Lambda}{3} \right] \label{HC}
\end{equation}
where $\Tilde{N}:= \bar N'$ is the new lapse function. The term $2GM$ can be introduced either by imposing appropriate boundary conditions to ensure the existence of Schwarzschild-like solutions \cite{Kuchar1994,CGP2007,GP2013} in the limit of vanishing cosmological constant, or by noticing that it is actually a Dirac observable as shown in \cite{Kuchar1994,oliveira}. The Hamiltonian constraint now has an Abelian algebra with itself and the usual algebra with the diffeomorphism constraint,
\begin{align}
     \{H_{\rm ab}(\Tilde{N}), H_{\rm ab}(\Tilde{M}) \} &= 0,\\
     \{H(\Tilde{N}),H_x(\bar N_x)\} &= H_{\rm ab}(\bar N_x \Tilde{N}'),\\
     \{H_x(\bar N_x), H_x(\bar M_x)\} &= H_x(\bar N_x \bar M'_x - \bar N'_x \bar M_x).
\end{align}
Classically, as it was shown in \cite{GP2013} for the Schwarzschild spacetime, the Hamiltonian constraint \eqref{HC} can be factorized as $H_{\rm ab}(\Tilde{N})=\int dx \Tilde{N} H_{-}H_{+}$ where
\begin{eqnarray}
    H_{\pm} = \sqrt{\sqrt{|E^x|}(1+ K^2_{\varphi})- 2GM - |E^x|^{3/2} \frac{\Lambda}{3} } \pm \frac{(E^x)'(|E^x|)^{1/4}}{2|E^{\varphi}|}.
\end{eqnarray}
Classically $(E^x)'$ is positive definite, hence the vanishing of the Hamiltonian constraint can then be taken to correspond to $H_{-}=0$. This simplification leads to a first order differential equation in the quantum theory for the physical states (see Appendix \ref{app:qtheory}). Accordingly, we redefine the lapse to be $\underline{N} = \Tilde{N} H_{+}/2E^\varphi$ to rewrite the Hamiltonian constraint in the form
\begin{equation}
    H_{\rm ab}(\underline N) = \int  dx \underline N \left( 2 |E^\varphi| \sqrt{\sqrt{|E^x|}(1+ K^2_{\varphi})- 2GM - |E^x|^{3/2} \frac{\Lambda}{3} } - (E^x)'|E^x|^{1/4} \right).
\end{equation}

Classically, the constraints can be easily solved. For simplicity, we restrict to the stationary slicing which corresponds to $\bar N^x=0$ and $\tilde N=0$, yielding $N = a (E^x)'/|E^\varphi|$ and $N^x = -2N K_\varphi \sqrt{|E^x|}/(E^x)'$, where $a$ is a constant of integration. The two constraints reduce two degrees of freedom per spacetime point,  leaving two residual gauge degrees of freedom. In order to remove them and fully gauge-fix the theory, we use two functional parameters to express two of the phase space variables as
\begin{equation}
    E^x = g(x), \quad K_\varphi = h(x),
\end{equation} 
which may actually depend on $M$ and $\Lambda$ as well. With these choices, the theory can be consistently solved to yield the remaining phase space variables as
\begin{eqnarray}
    (E^\varphi (x))^2 &=& \frac{g'(x)^2 /4}{1+h^2(x) - \frac{2GM}{\sqrt{g(x)}}- \frac{\Lambda }{3} g(x)}, \label{Ephi2} \\
    K_x(x) &=& \frac{h'(x)/2}{\sqrt{1+h^2(x) - \frac{2GM}{\sqrt{g(x)}}- \frac{\Lambda }{3} g(x)}},
\end{eqnarray}
where we require that $g(x)>0$ and $g'(x) \neq 0$. Different choices of the arbitrary functions $h(x)$ and $g(x)$ correspond to different coordinate choices for the stationary spacetimes represented by our theory. In order to fix the constant of integration in the lapse function, we further require that the resulting spacetimes in the limit of vanishing cosmological constant to be asymptotically flat \cite{Kuchar1994,CGP2007,GP2013}. We then must have $g(x)=x^2 + O(x^{-1})$ and $h(x)=O(x^{-1})$ when $x \rightarrow \infty$. This determines the integration constant in $N$ to be $a=1/2$. Thus we finally have
\begin{eqnarray}
    N^2&=&1+h^2(x) - \frac{2GM}{\sqrt{g(x)}}- \frac{\Lambda }{3} g(x), \label{lapse} \\
    N^x &=& -\frac{2 h(x) \sqrt{g(x)}}{g'(x)} \sqrt{1+h^2(x) - \frac{2GM}{\sqrt{g(x)}}- \frac{\Lambda }{3} g(x)} \label{shift}.
\end{eqnarray}

\section{Quantum theory: kinematics of the improved dynamics scheme}
\label{SecIII}
The kinematical Hilbert space of loop quantum gravity is the space of cylindrical functions of the holonomies of the connections defined over arbitrary graphs, where the spin network states provide an orthonormal basis for this space \cite{RovelliBook,ThiemannBook,StatusReport}. For our reduced spherically symmetric model, the details of these spin network states and the kinematical Hilbert space are extensively discussed in previous works \cite{B2004,BS2006,CGP2007,GP2013,GOP2014}. The basic elements are one-dimensional oriented graphs $g$ with support in the radial direction, which are composed by edges $\{e_j\}$ along the radial direction connecting the vertices $\{v_j\}$. The connection component $K_x$ is associated with holonomies in the radial direction and $K_\varphi$ with point holonomies on the vertices. We follow the construction suggested in \cite{GOP2023,BMO2024} and consider spin networks with a finite but arbitrarily large number of edges and vertices. The kinematical Hilbert space of the reduced model has a basis of spin network states $|\Vec{k},\Vec{\mu}\rangle$, with $k_j \in \mathbb{Z} $ the valences of the edges $e_j$ and $\mu_j \in \mathbb{R} $ the valences of the vertices $v_j$. The action of the triad operators on this basis is given by
\begin{eqnarray}
    \hat{E}^x (x_j) |\Vec{k},\Vec{\mu}\rangle &=& \ell^2_{Pl} k_{j} |\Vec{k},\Vec{\mu}\rangle \quad \text{if} \quad x_j \in e_j, \label{Ex}\\
    \hat{E}^\varphi (x) |\Vec{k},\Vec{\mu}\rangle &=& \sum_{v_j} \delta(x-x_j) \ell^2_{Pl} \mu_j |\Vec{k},\Vec{\mu}\rangle, \label{Ephi}
\end{eqnarray}
It is worth commenting that $(\ell^2_{Pl} k_j)$, the eigenvalues of $\hat{E}^x (x)$, can be naturally identified with areas of the spheres of symmetry normalized by the (dimensionless) unit sphere area. However, the eigenvalues of $\hat{E}^\varphi (x)$ have no invariant geometrical meaning since the classical analog is a scalar density (it is not invariant under coordinate transformations). 

The variables conjugate to the triads in the quantum theory are the holonomies of the connections along the edges of the graph. The Abelianized Hamiltonian constraint \eqref{HC} only contains the connection component $K_\varphi$, which is represented in the quantum theory by point holonomies $\hat{\mathcal{U}}_{\rho_j}:=\widehat{{\rm exp}(i\rho_j K_\varphi (x_j))}$ which act on the vertices. Their action on the basis states is given as 
\begin{equation}
    \hat{\mathcal{U}}_{\rho_j} (x_j) |\Vec{k},\Vec{\mu}\rangle = |\Vec{k},\Vec{\mu'}\rangle,
\end{equation}
where $\Vec{\mu'}$ is obtained from $\Vec{\mu}$ by modifying the valence $\mu_j$ of the vertex located at $x_j$ by $\mu_j + \rho_j$. For the mass $\hat M$ we adopt a standard representation.

Before we proceed with the quantization program, it is worth introducing here the improved dynamics scheme following the ideas in \cite{Chiou2012} and \cite{dBOP2017,GOP2020,BMO2024}. It allows us to fix various deformation parameters that appear in the theory, while they respect the simplicity of the above kinematics which will still be suitable for further calculations. The motivation is that, classically, the curvature components at a point are equivalent to the holonomies of the connection around infinitesimal closed loops along suitable directions. However in LQG, various geometric operators such as area and volume have discrete spectra. Thus, the curvature components in LQG are approximated by holonomies around plaquettes which enclose the smallest (finite) possible area allowed by the theory, which in this case we choose to equal the minimum eigenvalue of the area operator in LQG, denoted by $\Delta$. In this reduced model we choose the plaquettes to be well adapted to the Killing symmetries which, once equated to the minimum area eigenvalue, yielding relations which fix the kinematical parameters.

Concretely, let us first consider the plaquettes adapted to the 2-spheres at each vertex. Classically the 2-spheres have a physical area given by $4\pi g_{\theta \theta}(x) = 4\pi E^x (x)$. In order to impose the improved dynamics prescription, we assume that the areas of the 2-spheres in the effective geometry obtained from the quantum dynamics are well approximated by replacing $E^x$ in the above expression by its eigenvalue given in \eqref{Ex}. The validity of this assumption rests on whether we obtain a self-consistent and physically sensible quantization at the end of the day. The plaquette on the $\theta-\phi$ sector will hence enclose an area given by $4\pi \ell^2_{Pl} |k_j|\rho_j^2 $ which must be equal to the minimum area eigenvalue, $\Delta$, which yields
\begin{equation}
\label{rho_j}
    \rho_j = \sqrt{\frac{\Delta}{4\pi \ell^2_{Pl} |k_j|}}.
\end{equation} 
This suggests a more convenient state relabeling as $|\nu_j\rangle$ where $\nu_j=\mu_j \sqrt{4\pi \ell^2_{Pl} |k_j|/\Delta}$. This simplifies the action of the point holonomies as
\begin{equation}
    \hat{\mathcal{U}}_{\rho_j}|\nu_j\rangle = |\nu_j +1\rangle.
\end{equation}
The physical meaning of this label was discussed in \cite{dBOP2017} and can be naturally identified to be proportional to the local volume operator $\hat V(x_j)=\sqrt{|\hat E^x(x_j)|}\hat E^\varphi (x_j)$.

Further, as introduced in \cite{BMO2024}, we implement a second improved dynamics condition on the plaquettes in the $\theta-x$ and $\varphi-x$ planes, which lead to only one additional condition due to spherical symmetry. We set these plaquettes in the equatorial plane ($\theta = \pi/2$) without loss of generality. Classically, infinitesimal lengths along the $x$ and $\varphi$  directions are equal to the norms of the $1$-forms $(dx)^\mu$ and $(d\phi)^\mu$ respectively. Thus the area of an infinitesimal plaquette in the equatorial plane can be written as $(\sqrt{|g_{\mu \nu}|(dx)^\mu(dx)^\nu}\sqrt{|g_{\mu \nu}|(d\varphi)^\mu(d\varphi)^\nu})$. This expression for the area is coordinate independent, so we can simplify it by going to the diagonal gauge in which the metric is diagonal (this amounts to setting $h(x)=0$), leading to the expression $(\sqrt{|g_{x x}|}dx\sqrt{|g_{\theta \theta}|}d\varphi)$ for the area in the diagonal gauge. This leads to the improved dynamics condition
\begin{equation}
\label{condition2a}
    2\pi \sqrt{|g_{x x}(x_j)|}\delta x_j \sqrt{|g_{\theta \theta} (x_j)|} \rho_j = \Delta,
\end{equation}
at the vertex $v_j$, where $2\pi \rho_j$ is the coordinate length along the $\varphi$ - direction and $\delta x_j$ is the coordinate length along the $x$ - direction, namely, of the edge $e_j$ of the spin network as per our kinematical scheme described above. For simplicity, we will not impose this condition at all vertices, but only in the vertex where we expect quantum corrections will be largest. This suggests a choice of spin networks that show an equal spacing in a suitable radial coordinate, such that $\delta x_j = \delta x$. In particular, we will restrict our analysis, without loss of generality, to spin networks such that 
\begin{equation}
    \hat{E}^x(x) |\Vec{k},\Vec{\nu}\rangle = \ell^2_{Pl}k_j |\Vec{k},\Vec{\nu}\rangle = \text{sign}(x_j)(x_j^2 + x_0^2)|\Vec{k},\Vec{\nu}\rangle,
\end{equation}
where
\begin{eqnarray}
    x_j&=&j\delta x \quad \text{if} \quad j\in [-S,-1], \\
    x_j&=&(j-1)\delta x \quad \text{if} \quad j\in [1,S],
\end{eqnarray}
such that $(\delta x/ \ell_{Pl}) \in \mathbb{N}$. This construction follows from \cite{GOP2023,BMO2024}.

The required classical metric components in the diagonal gauge for the above improved dynamics conditions are given by
\begin{equation}
    g_{\theta \theta}(x) = E^x (x), \quad g_{x x}(x) = \frac{(E^\varphi (x))^2}{|E^x (x)|} = \frac{([E^x (x)]')^2}{4E^x(x)}\frac{1}{1-\frac{2GM}{\sqrt{E^x (x)}}-\frac{\Lambda}{3}E^x(x)},
\end{equation}
where we have used equation \eqref{Ephi2} in the last expression. Following \cite{BMO2024}, we approximate $([E^x (x)]')^2$ by $(2\sqrt{x_j^2+ \Delta^2/x_0^2}+\delta x)^2$, which agrees with the exact expression up to corrections of the order $\Delta^2/x_0^2$, which will be negligible for macroscopic black holes, as we will see. With these choices, the improved dynamics condition \eqref{condition2a} evaluated at $j=1$, reduces to 
\begin{equation}
\label{condition2b}
    \frac{\delta x}{2 x_0} \frac{(2\Delta/x_0+\delta x)}{\sqrt{\frac{2GM}{x_0}+\frac{\Lambda x_0^2}{3}-1}} = \sqrt{\frac{\Delta}{\pi}},
\end{equation}
where we have used the first improved dynamics condition \eqref{rho_j} to substitute for $\rho_j$. Hence, this is another condition which fixes the deformation parameter $\delta x$ in our quantum theory.

\section{Physical states and observables}

The quantum dynamics of the model is summarized in Appendix \ref{app:qtheory}. On physical states, the kinematical operators like  $(\hat E^x)$ and $(\hat E^\varphi)^2$ are written as parameterized observables. This implies they can be expressed in terms of both Dirac observables and functional parameters, the latter possibly being also functions of Dirac observables. Concretely, 
\begin{eqnarray}
&&    \hat E^x(x) = \hat O_{z(x)},\quad \widehat{ [E^x(x)]' }= \frac{\hat O_{z(x)+1/S}- \hat O_{z(x)}}{x(z+1/S)-x(z)},\\
&&    (\hat{E}^\varphi(x_j))^2= \frac{([\hat{E}^x(x_j)]')^2 /4}{1+\frac{\sin^2(\widehat{\rho_j K_\varphi} (x_j))}{\rho^2_j} - \frac{2G\hat{M}}{\sqrt{|\hat{E}^x(x_j)|}}- \frac{\Lambda }{3} |\hat{E}^x(x_j)|}, \label{Ephi2_operator}
\end{eqnarray}
where $2S$ is the total number of vertices. Besides, $\hat M$ and $ \hat O_j$ are Dirac observables, the latter with spectrum given by $(\ell_{Pl}^2k_j)$,\footnote{These $ \hat O_j$ are promoted to the parametrized observables $ \hat O_{z(x)}$ once we define $j$ in terms of the parametrized function of $z(x)\in[-1,1]$ by means of $j(z) = {\rm Int}(Sz)$.} while $z(x)\in[-1,1]$ and $K_{\varphi}(x)$ are, from now on, functional parameters, namely, prescribed functions that may also depend on Dirac observables. In order for $(\hat{E}^\varphi)^2$ to be a well-defined self-adjoint operator, it must satisfy $(\hat{E}^\varphi(x_j))^2 >0$. From the above expression, we see that there are two ways that $(\hat{E}^\varphi(x_j))^2$ may fail to be positive-definite - either when the cosmological constant is positive and the product $\Lambda |\hat{E}^x|$ becomes too large, or when $|\hat{E}^x|$ is too small. We will see that this leads to an upper bound for $\Lambda$ and a lower bound for $|\hat{E}^x|$, respectively, in order for $(\hat{E}^\varphi(x_j))^2$ to be well-defined. To understand this, we identify the source of the problem in different regimes. 

Let us first focus on the asymptotic region where $|\hat{E}^x|$ is macroscopic and large, i.e. $|\hat{E}^x| \approx x_j^2$ where $j$ is so large that we may approximate it by a continuous label. In this regime, the improved dynamics condition \eqref{rho_j} implies that $\rho_j$ is vanishingly small. For choices of the gauge function $K_\varphi(x_j)$ such that $\sin^2(\rho_j K_\varphi(x_j)) \simeq \rho_j^2 K^2_\varphi(x_j)\simeq {\cal O}(x_j^{-\alpha})$ with $\alpha\geq 0$, the quantum expression \eqref{Ephi2_operator} for $(\hat{E}^\varphi(x_j))^2$ reduces to the classical expression \eqref{Ephi2}, as expected. However, when $\Lambda$ is positive, the classical expression \eqref{Ephi2} for $[E^\varphi(x_j)]^2$ itself may be ill-defined when $E^x (x)$ exceeds a certain maximum value (say at $E^x(x=x_L)$, which depends on $\Lambda$, $M$ and the choice of gauge function $h(x)$ provided it is chosen such that it does not grow as $x^2$). If $x_L$ is large enough, one may be inclined to think that we are far from the strong quantum region at the center, and classical expressions should faithfully approximate quantum dynamics in any valid gauge choice. Consequently, we may conclude that we need not view the failure of $[E^\varphi(x_j)]^2$ to be positive-definite as a problem with the quantization, or putting a limit on how large a positive $\Lambda$ can be, as we are already at the classical level where this can be easily addressed by making a different gauge choice for $h(x)$ where the slicing will be valid beyond $x_L$, as shown for example in the next section with Eddington-Finkelstein coordinates where this problem disappears. However, as we find out in section VB that it turns out to be more than a mere failure of the choice of slicing. In particular, we find that the curvature of spacetime in the asymptotic limit is of the order of $\Lambda$. The curvature there may be arbitrarily large if $\Lambda$ is allowed to be arbitrarily large. In particular, we expect quantum effects to become relevant when the curvature is positive and approaches Planckian values. This can be seen as follows. In order for $(\hat{E}^\varphi)^2$ to be well-defined, the denominator in equation \eqref{Ephi2_operator} must be positive-definite. The best we can do at the quantum level is to choose $\rho_j K_\varphi(x_j) = \pi/2$, implying that the two positive terms in the denominator of \eqref{Ephi2_operator} reach a global maximum given by $1+1/\rho_j^2$, independent of the choice of gauge (but is manifest for the choice made few line above). This leads to the condition
\begin{equation}
\label{condition3a}
    1+\frac{4\pi (x_j^2+x_0^2)}{\Delta} - \frac{2GM}{\sqrt{x_j^2+x_0^2}}- \frac{\Lambda }{3} (x_j^2+x_0^2) > 0, \quad \forall \quad  x_j,M,\Lambda,
\end{equation}
where we have substituted for $\rho_j$ from \eqref{rho_j}. In order to obtain nontrivial semiclassical spacetimes, the choice of $\Lambda$ must allow for an arbitrary number of vertices such that we can take the limit $\ell_{Pl}^2k_j = x_j^2 +x_0^2\to +\infty$. In the asymptotic limit $x_j \to \infty$, the above condition leads to a mass-independent Planckian upper bound on the cosmological constant given by
\begin{equation}
\label{Lambda_max}
    \Lambda < \Lambda_{max} = \frac{12\pi}{\Delta}.
\end{equation}
In other words, the cosmological constant must satisfy $-\infty<\Lambda<\Lambda_{max}$. It is an interesting question whether a quantization of the Schwarzschild-de Sitter spacetime consistent with $\Lambda$ greater than this threshold can be obtained. In the case of homogeneous de Sitter cosmologies \cite{posLambda-LQC,posLambda-LQC2}, the spectrum of the Hamiltonian constraint operator was studied, and it was not possible to find normalizable physical states for a positive $\Lambda$ larger than a critical (Planck order) value. Hence, the quantum theory provided a nontrivial bound $\rho_\Lambda < \rho_c = 3 / \Delta$. Here, we find a similar result but for spherically symmetric spacetimes. As we shall see in the next section, the curvature invariants in the asymptotic limit are proportional to $\Lambda$. Thus, this Planckian upper bound on $\Lambda$ coming form the quantization serves to ensure that the curvature cannot exceed a certain positive Planckian upper bound in the asymptotic limit. 

Note that our quantization does not lead to an upper bound on the magnitude of $\Lambda$ when it is negative. In other words, for the Schwarzschild-anti-de Sitter spacetimes for which the curvature in the asymptotic limit is again proportional to $\Lambda$ and hence negative, our quantization in principle allows to choose a negative $\Lambda$ larger than Planckian values which makes the negative curvature at conformal infinity larger than Planckian. This is of little phenomenological interest as the observed values of $\Lambda$ are insignificantly small. However, conceptually, it seems an intriguing  problem worth exploring. Note that unlike the de Sitter cosmology, loop quantization of the homogeneous anti-de Sitter spacetime also does not lead to any upper bound on the negative $\Lambda$ \cite{AdS_LQC}. These results seem to suggest that quantum gravity does not limit the magnitude of a finite negative curvature (a repulsive gravitational field), however large. Intuitively, a repulsive gravitational field does not lead to the kind of runaway catastrophic phenomenon (such as gravitational collapse) that an attractive field does if it is too large. Therefore, this may explain why no limit is obtained on how large a negative $\Lambda$ can be.

We must also look at the consequences of the condition \eqref{condition3a} in the regime where $x_j \to 0$. This is the quantum regime when the holonomies are nearly saturated and $\hat{E}^x = x_0^2$ (corresponding to the minimum area $2$-spheres). This leads to the condition 
\begin{equation}
    1+\left( \frac{4\pi}{\Delta} - \frac{\Lambda }{3} \right)x^2_0 - \frac{2GM}{x_0} >0 \quad \forall \quad  M,\Lambda.
\end{equation}
Given $M$ and $-\infty<\Lambda<\Lambda_{max}$, this leads to a lower bound on $x_0$, which can be found as follows. We look at the roots of the following expression
\begin{equation}
\label{condition3b}
    \sigma = 1+\left( \frac{4\pi}{\Delta} - \frac{\Lambda }{3} \right)x^2_0 - \frac{2GM}{x_0},
\end{equation}
which are the same as the roots of the polynomial $x_0-2GM + A x_0^3$ where $A=( 4\pi / \Delta - \Lambda /3 )$. Since $\Lambda < 12\pi / \Delta$ as per the condition \eqref{Lambda_max}, $A$ is positive definite and the discriminant of the above cubic polynomial is negative definite, implying the existence of a single real root, which provides a lower bound for $x_0$ below which the expression $\sigma$ is negative and the condition \eqref{condition3b} is violated. In the limit $2GM \gg \ell_{Pl}$ and $|\Lambda| \ll 12\pi / \Delta$, the lower bound on $x_0$ is given by
\begin{equation}
\label{xmin}
    x_0^{min} \approx \left( \frac{2GM}{4\pi/\Delta-\Lambda/3} \right)^{1/3}.
\end{equation}
Thus the condition \eqref{condition3a} has led us to the conclusion that we must only consider spin networks which have support on eigenvalues of the operator $\hat{E}^x(x)$ greater than $(x^{min}_0)^2$. Moreover, in the limit $|\Lambda| \ll 12\pi / \Delta$, $2GM \gg \ell_{Pl}$, the second improved dynamics condition implies 
\begin{equation}
\label{delta_x}
    \delta x \approx 2 \ell_{Pl} \text{Int} \left[ \frac{x^{min}_0}{\ell_{Pl}} \right],
\end{equation}
at leading order. Thus we see that the two improved dynamics conditions \eqref{rho_j} and \eqref{condition2b} and the condition \eqref{condition3a} help us to fix the three kinematical parameters $x_0$, $\rho_0$ and $\delta x$.

As shown in \cite{BMO2024}, the leading order term in the value obtained for $x^{min}_0$, which is $(2GM \Delta/ 4\pi)^{1/3}$ is the same as that for the charged black hole analyzed in \cite{BMO2024} as well as the uncharged Schwarzschild black hole studied in \cite{GOP2020,GOP2021} in the improved dynamics scheme. The corrections to the leading order term produced at the first order by the presence of a positive cosmological constant are positive, i.e. $x^{min}_0$ is larger here, in contrast to the charged black hole considered in \cite{BMO2024}, where the first order corrections due to presence of a tiny amount of charge are negative, making $x^{min}_0$ smaller. \footnote{For a solar mass black hole, $x_0^{min}\simeq 10^{-23}m \simeq 10^{12} \ell_{Pl}$ at the leading order. For supermassive black holes with masses about a billion times that of the sun $x_0^{min} \simeq 10^{-20}m$. Thus $x_0^{min}$ would be microscopic even for the largest known black holes.}  On the other hand, if the cosmological constant is negative, $x_0^{min}$ becomes smaller. If $\Lambda<0$ and in the limit $|\Lambda|\gg 2GM \ell_{Pl}$, $x_0^{min}\simeq \ell_{Pl}$ since the quantum theory does not allow $x_0^{min}=0$, otherwise the parametrized observable $(\hat E^\varphi)^2$ would be ill defined. Moreover, in the case of $\Lambda<0$, the location of the cosmological horizon decreases as $\Lambda^{-1/3}$. Again, in the limit $|\Lambda|\gg 2GM \ell_{Pl}$, its value can be as small as  $\ell_{Pl}$, but not smaller, for similar reasons already explained. We will discuss this in more detail in next section. 

\section{Effective Geometry}
In this section, we obtain the effective metric from the above quantization in the improved dynamics, which will help us study its physical aspects. The effective metric is also crucial in analyzing various phenomenological properties of the quantized spacetime. In order to compare with previous quantizations of the Schwarzschild black hole \cite{GOP2020,GOP2021} and the charged black hole \cite{BMO2024} using improved dynamics, we use the same slicing in this section as used in these studies, which leads to Eddington-Finkelstein coordinates. As we will see, this will also alleviate the difficulties in analyzing the asymptotic region in the presence of the cosmological constant as commented in the previous section. The Eddington-Finkelstein horizon penetrating coordinates amount to the following gauge fixing \cite{GOP2020,GOP2021}
\begin{equation}
    \frac{\sin^2(\widehat{\rho_j K_\varphi} (x_j))}{\rho^2_j} = \frac{\left(\frac{ 2G\hat{M}}{\sqrt{|\hat{E}^x(x_j)|}}+ \frac{\Lambda }{3} |\hat{E}^x(x_j)| \right)^2}{1 + \frac{2G\hat{M}}{\sqrt{|\hat{E}^x(x_j)|}}+ \frac{\Lambda }{3} |\hat{E}^x(x_j)|}.
\end{equation}
Equations \eqref{lapse} and \eqref{shift} immediately let us obtain the operator expressions for lapse and shift in this case. Using equation \eqref{Ephi2}, we find that
\begin{equation}
    (\hat{E}^\varphi (x_j))^2= \frac{([\hat{E}^x(x_j)]')^2}{4} \left(
    1 + \frac{2G\hat{M}}{\sqrt{|\hat{E}^x(x_j)|}}+ \frac{\Lambda }{3} |\hat{E}^x(x_j)| \right),
\end{equation}
which is positive definite in the Eddington-Finkelstein coordinates. Treating the metric components given in \eqref{metric} as parameterized observables, their operators are given by
\begin{eqnarray}
    \hat{g}_{xx} (x_j) &=& \frac{(\hat{E}^\varphi (x_j))^2}{|\hat{E}^x(x_j)|} = \frac{([\hat{E}^x(x_j)]')^2}{4 |\hat{E}^x(x_j)|} \left(
    1 + \frac{2G\hat{M}}{\sqrt{|\hat{E}^x(x_j)|}}+ \frac{\Lambda }{3} |\hat{E}^x(x_j)| \right), \\
    \hat{g}_{tx} (x_j) &=& \hat{g}_{xt} (x_j) =  N_x = g_{xx} N^x = -\frac{[\hat{E}^x(x_j)]'}{2 \sqrt{|\hat{E}^x(x_j)|}} \left(\frac{ 2G\hat{M}}{\sqrt{|\hat{E}^x(x_j)|}}+ \frac{\Lambda }{3} |\hat{E}^x(x_j)| \right), \\
    \hat{g}_{tt} (x_j) &=& -(N^2 - N_x N^x) = - \left(
    1 - \frac{2G\hat{M}}{\sqrt{|\hat{E}^x(x_j)|}} - \frac{\Lambda }{3} |\hat{E}^x(x_j)| \right), \\
    \hat{g}_{\theta \theta}(x_j) &=& |\hat{E}^x(x_j)|, \quad \hat{g}_{\varphi \varphi}(x_j) =  |\hat{E}^x(x_j)| \sin^2 \theta.
\end{eqnarray}
To obtain the effective metric, we follow the strategy used in \cite{GOP2020}. We restrict to a single spin network and consider a family of states sharply peaked at large values $M$ and choosing $-\infty<\Lambda<\Lambda_{max}$. The details of the construction are analogous to those of \cite{GOP2020}. An effective spacetime can then be defined as the one obtained by computing the expectation values of the metric operators on these sharply peaked states, i.e. $g_{\mu \nu} = \langle \hat{g}_{\mu \nu} \rangle$. While we have already neglected the effects emerging from superpositions of several spin network states, this effective metric still inherits several quantum corrections from the underlying quantum theory, namely, (i) the lower bound \eqref{xmin} on the value of $\hat{E}^x(x)$, (ii) corrections due to polymerization of the curvature components, (iii) the inherent discreteness in derivatives such as $[\hat{E}^x(x_j)]'$ and the derivatives of the metric needed to compute curvature components, and (iv) superpositions in $\hat{M}$. However, here we focus only on the most prominent effects which are due to (i) and (ii). We ignore the subleading contributions due to the spread $\Delta M$ in mass. These effects are discussed in \cite{GOP2020}. And for simplicity, we ignore the effects of discreteness of the spectrum of $\hat{E}^x(x)$ and the discreteness of all the derivative terms, emerging from the fact that we have a finite number of discrete vertices. We assume we are working at a certain coarse-grained level where we can replace $x_j$ with a continuous variable $x$. The justification for this simplification is also provided in \cite{GOP2020}, where numerical values of the discrete and coarse-grained expressions of the second derivatives of the metric are compared to show the error incurred in coarse-graining is at most $10\%$ in the most quantum region, which quickly becomes negligible as we move to low curvature regions. Under these assumptions, the effective metric can be written as $g_{\mu \nu} = {}^{(0)} g_{\mu \nu} + ....$, where ${}^{(0)} g_{\mu \nu}$ contains only the corrections due to effects (i) and (ii) and is given by
\begin{eqnarray}
    ^{(0)}ds^2: &=& ^{(0)} g_{\mu \nu} dx^{\mu} dx^{\nu} = -f(x)dt^2 - 2(1-f(x))\frac{\left(\sqrt{x^2+ \Delta^2/(x_0^{min})^2}+ x_0^{min}\right)}{\sqrt{x^2+ (x_0^{min})^2}}dtdx \nonumber \\
    && + (2-f(x))\frac{\left(\sqrt{x^2+ \Delta^2/(x_0^{min})^2}+ x_0^{min}\right)^2}{x^2+ (x_0^{min})^2}dx^2 + \left( x^2+ (x_0^{min})^2 \right) d\Omega^2, \label{effectivemetric}
\end{eqnarray}
where
\begin{equation}
    f(x) = 1 - \frac{2GM}{\sqrt{x^2+ (x_0^{min})^2}} - \frac{\Lambda}{3} \left( x^2+ (x_0^{min})^2 \right),
\end{equation}
where we have used \eqref{delta_x} to replace $\delta x$.

\subsection{Causal structure of the effective spacetime}
As we investigate the causal structure, we find that there are four distinct spacetimes depending on the value of the cosmological constant. We describe the causal structure of the effective spacetime for all four possibilities. 

Firstly, to locate if there are any horizons, we consider the Killing vector field $t^\mu$ that in the case $\Lambda = 0$ agrees with static observers with respect to the black hole and is time-like at spatial infinity. Its norm is given by $t^\mu t_\mu = -f(x)$, indicating that the roots of the equation $f(x)=0$ correspond to surfaces $x=$ const on which the Killing vector field becomes null. Thus, the location of horizons is given by
\begin{eqnarray}
    1 - \frac{2GM}{\sqrt{x^2+ (x_0^{min})^2}} - \frac{\Lambda}{3} \left( x^2+ (x_0^{min})^2 \right) = 0.
\end{eqnarray}
To solve the above equation, we make the substitution
\begin{eqnarray}
\label{r}
r^2= x^2+ (x_0^{min})^2,\quad x>0,\quad r>0.
\end{eqnarray}
Here we consider only the positive square root of $x^2+ (x_0^{min})^2$, i.e. $r>0$, as the other choice does not yield a Schwarzchild spacetime. With this substitution, the above condition reduces to the corresponding condition for the classical Schwarzschild-de Sitter spacetime given by
\begin{eqnarray}
\label{horizons}\label{eq:roots}
    1 - \frac{2GM}{r} - \frac{\Lambda}{3} r^2 = 0,
\end{eqnarray}
whose roots are well known. The nature of the roots of the equation \eqref{horizons} depends on the value of $\Lambda$. We now describe all the possibilities one by one.

\subsubsection{Schwarzschild-de Sitter spacetime with $0<9G^2 M^2\Lambda<1$}

When $0<\Lambda<1/9G^2 M^2$ (small spacetime masses), Eq. \eqref{eq:roots} has three real roots, two of which are positive (and physical since we choose $r>0$) given by
\begin{equation}
    r_{\pm} = -\frac{2}{\sqrt{\Lambda}} \cos \left( \frac{1}{3} \cos^{-1}(3GM\sqrt{\Lambda}) \pm \frac{2\pi}{3} \right). 
\end{equation}
Here $r_{+}>r_{-}>0$. Classically, there is a cosmological horizon at $r_{+}$ and a black hole horizon at $r_{-}$. Using these and equation \eqref{r}, we can easily obtain the horizons for the effective geometry as follows
\begin{equation}
    x_\pm=\sqrt{r_{\pm}^2 -(x_0^{min})^2}.
\end{equation}
These two positive roots $\sqrt{r_{\pm}^2 -(x_0^{min})^2}$ correspond to the cosmological and black hole horizons just as in the classical Schwarzschild-de Sitter spacetime. Note that the effective geometry is regular at $x=0$ and the spacetime can be extended to negative values of $x$. Thus we also have the roots corresponding to negative values of $x$. If we define $\tilde r^2= x^2+ (x_0^{min})^2$ for $x<0$ and with $\tilde r>0$, and inserting it in Eq. \eqref{eq:roots}, we obtain three real roots, two of them being positive (and physical), given by $\tilde x_{\pm}=-\sqrt{r_{\pm}^2 -(x_0^{min})^2}$. They correspond to the white hole horizon after the anti-trapped region in the effective spacetime ($\tilde x_{-}$) and the corresponding cosmological horizon ($\tilde x_{+}$).
%%%%%%%%%%%%%%%%%%%%%%%%%%%%%%%%%%%%%
\begin{figure}
    \centering
    \includegraphics[scale=0.5]{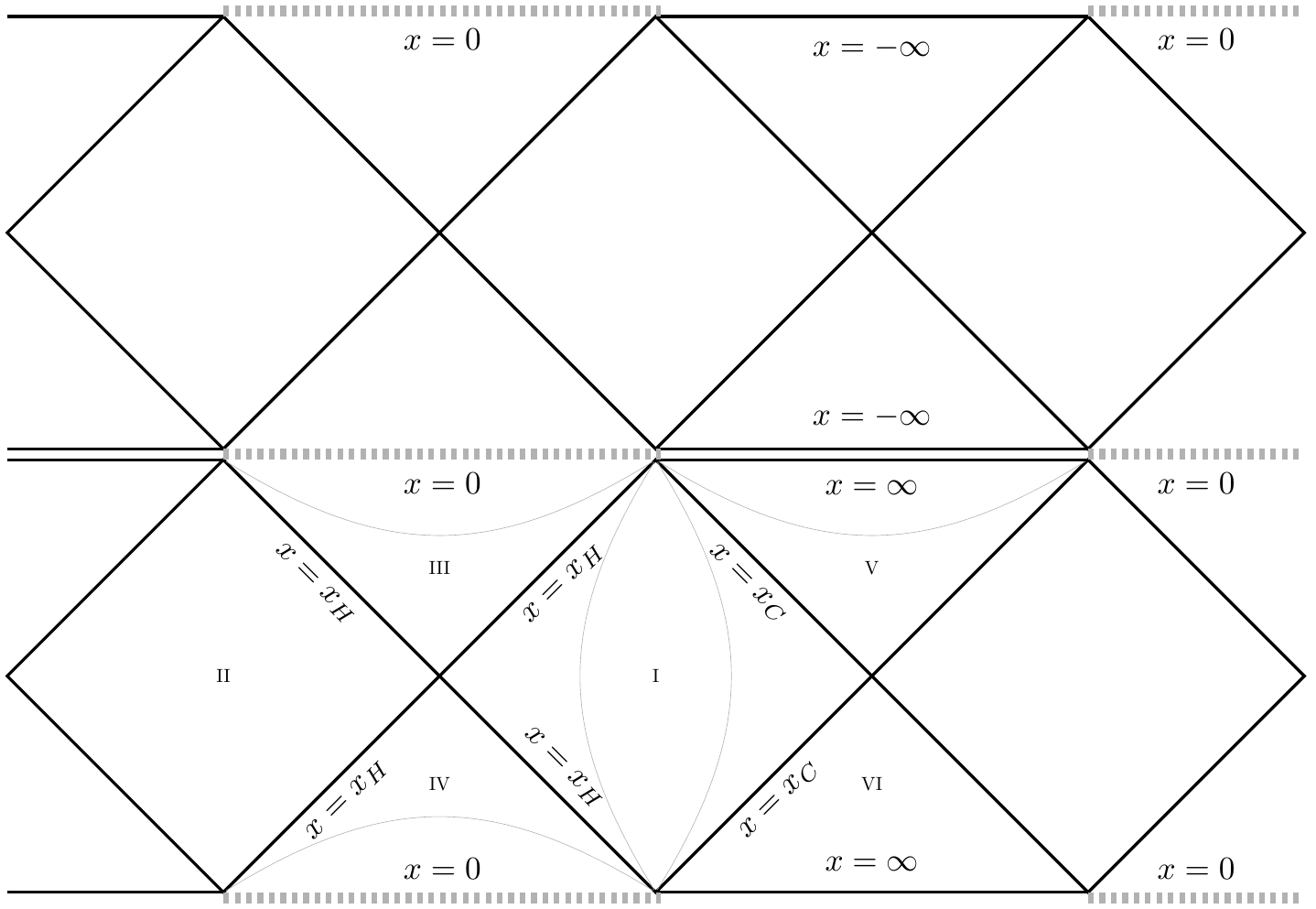}
    \caption{Penrose diagram for $\Lambda<1/9G^2 M^2$. Here, $x_H$ amounts to the location of the black/white hole horizons and $x_C$ to the location of the cosmological horizons. Horizontal dotted lines correspond to the transition surfaces. }
    \label{fig:conf-diagram1}
\end{figure}
%%%%%%%%%%%%%%%%%%%%%%%%%%%%%%%%%%%%%
 
In Figure \ref{fig:conf-diagram1}, we show the Penrose diagram for this configuration with $0<\Lambda<1/9G^2 M^2$. The main difference with the classical spacetime is that the singularity at $x=0$ is replaced by a transition surface connecting a trapped region (black hole interior) with an antitrapped region (white hole interior). Therefore, an observer falling into a black hole region will eventually pop up into another universe (after crossing the corresponding white hole horizon), and can either remain in the exterior region, fall into the corresponding black hole, or cross the cosmological horizon and reach the asymptotic infinity which now is a space-like surface. 

\subsubsection{Schwarzschild-de Sitter spacetime with $9G^2 M^2\Lambda>1$}

In the case of ultramassive spacetimes \cite{senovilla-23a,senovilla-23b}, namely, when $\Lambda>1/9G^2 M^2$, the Killing vector field $t^\mu$ is always space-like, while $\partial_x$ is always time-like and past directed. Hence, these spacetimes are homogeneous but anisotropic. In other words, the whole spacetime is free of horizons. It is conformed by a trapped region for $x>0$ and an anti-trapped region for $x<0$. Since the radial and time  coordinates become time-like and space-like, respectively, let us refer to the corresponding coordinates as $({\cal T},{\cal X})$. The line element, in the diagonal gauge, takes the form:
\begin{eqnarray}
    ^{(0)}ds^2: &=& -\frac{1}{f({\cal T})}\frac{\left(\sqrt{{\cal T}^2+ \Delta^2/{\cal T}_0^2}+ {\cal T}_0\right)^2}{{\cal T}^2+ {\cal T}_0^2}d{\cal T}^2 +f({\cal T})d{\cal X}^2+ \left( {\cal T}^2+ {\cal T}_0^2 \right) d\Omega^2, \label{effectivecosmo}
\end{eqnarray}
with
\begin{equation}
    f({\cal T}) = \frac{\Lambda}{3} \left( {\cal T}^2+ {\cal T}_0^2 \right)+\frac{2GM}{\sqrt{{\cal T}_0^2+ {\cal T}_0^2}}-1.
\end{equation}
\begin{figure}
    \centering
    \includegraphics[scale=1.0]{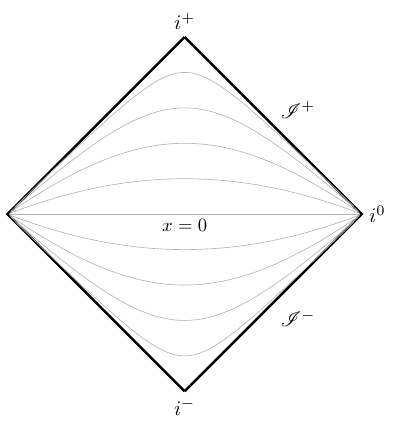}
    \caption{Penrose diagram for $\Lambda>1/9G^2 M^2$.}
    \label{fig:conf-diagram2}
\end{figure}

In Figure \ref{fig:conf-diagram2}, we show the corresponding Penrose diagram. The region ${\cal T}>0$ (i.e. $x>0$) corresponds to a trapped (collapsing) spacetime that reaches a Planckian curvature at ${\cal T}=0$ (i.e. $x=0$) and eventually transitions into an antitrapped (expanding) region ${\cal T}<0$ (i.e. $x<0$). Hence, this scenario can be understood as a  bouncing (anisotropic) cosmological model. We must keep in mind that the cosmological constant must still be bounded by Eq. \eqref{Lambda_max}. 

\subsubsection{Schwarzschild-de Sitter spacetime with $9G^2 M^2\Lambda=1$}

\begin{figure}
    \centering
    \includegraphics[scale=0.5]{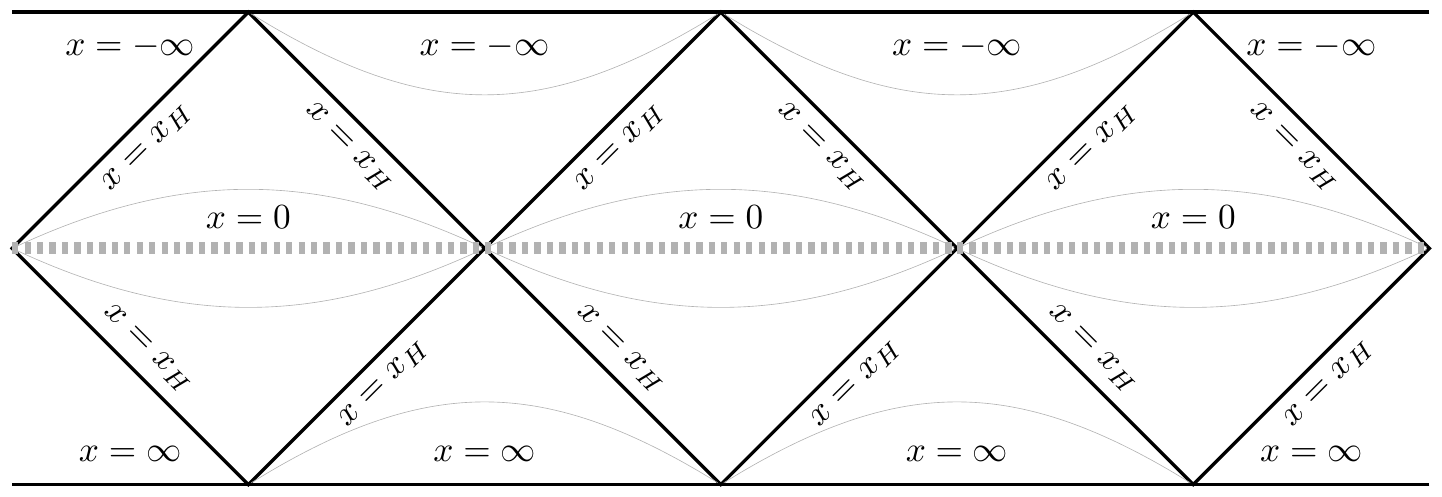}
    \caption{Penrose diagram for the extremal case $\Lambda=1/9G^2 M^2$. In this plot, $x_H$ amounts to the location of the black/white hole horizons and/or the cosmological horizons, since they both agree. Horizontal dotted lines correspond again to the transition surfaces. }
    \label{fig:conf-diagram3}
\end{figure}

Let us now discuss the particular case $\Lambda=1/9G^2 M^2$. Here, the two horizons given by \eqref{horizons} coincide when $\Lambda=1/9G^2 M^2$ and the spacetime only has one horizon. The Killing vector field $t^\mu$ is null at the horizon and space-like everywhere. The causal structure in this case is depicted in the Penrose diagram in Figure \ref{fig:conf-diagram3}. In this case observers emanate from the spatial infinity and cross the black hole horizon, entering a trapped region. Since the singularity at $x=0$ is resolved, the observers cross it entering an anti-trapped region and eventually come out of the white hole horizon, finally approaching the corresponding spatial infinity again.

\subsubsection{Schwarzschild-anti-de Sitter spacetime with $\Lambda<0$}
\begin{figure}
    \centering
    \includegraphics[scale=0.5]{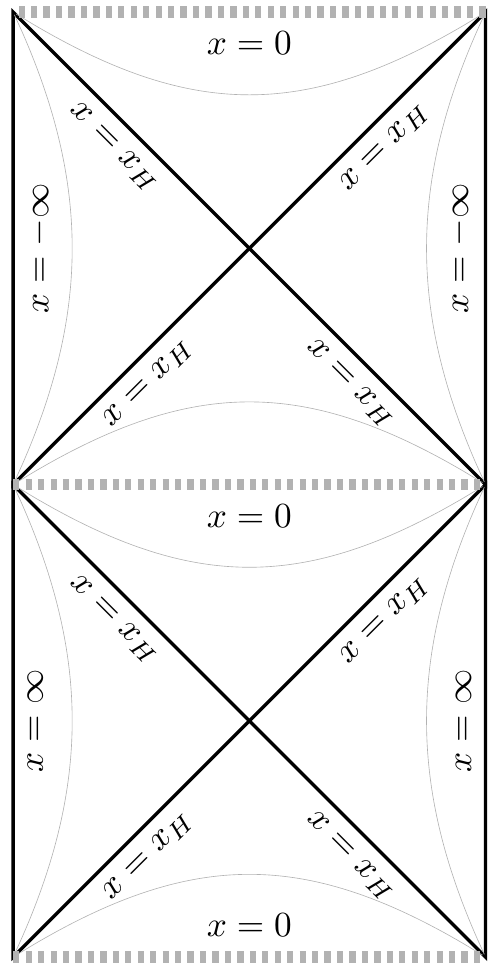}
    \caption{Penrose diagram for $\Lambda<0$. In this plot, $x_H$ amounts to the location of the black/white hole horizons. There are no cosmological horizons. Horizontal dotted lines correspond again to the transition surfaces.  }
    \label{fig:conf-diagram4}
\end{figure}

Finally, when $\Lambda<0$, the equation \eqref{horizons} has a single positive root given by
\begin{equation}
    r_h = \left(\frac{3GM}{\Lambda}\right)^{1/3} \left[(C+1)^{1/3} - (C-1)^{1/3} \right], \quad \text{where} \quad C=\sqrt{1-\frac{1}{9G^2 M^2 \Lambda}}.
\end{equation}
This single root $r_h$ corresponds to the black/white hole horizon. There is no cosmological horizon in this case. The causal structure in this case is depicted in Fig. \ref{fig:conf-diagram4}. The spacetime is asymptotically anti-de Sitter and the conformal infinity $x=\infty$ is time-like. It is interesting to note that, in the classical theory, where $x_0^{min}=0$, one can have $r_h\to 0$ if $\Lambda\to-\infty$. However, in the quantum theory, $x_0^{min}$ (and consequently $r_h$) cannot be less than the Planck length $\ell_{Pl}$. Thus, for a negative $\Lambda$ in the limit $|\Lambda|\gg 2GM \ell_{Pl}$, we have $x_0^{min}=\ell_{Pl}=r_h$, as this is the lowest value allowed for $x_0^{min}$ and $r_h$. While there is singularity resolution due to this lower bound on $x_0^{min}$, the curvature invariants can reach values larger than Planck scale in this limit. We discuss this in more detail below.

\subsection{Curvature of the effective spacetime}

We now use the effective metric \eqref{effectivemetric} to analyze the properties of the curvature invariants of the effective spacetime such as the Ricci scalar $R=R_{\mu\nu}g^{\mu\nu}$, the Kretschmann scalar $K=R_{\mu \nu \rho \lambda}R^{\mu\nu\rho\lambda}$ and the Ricci tensor squared $R_{\mu\nu}R^{\mu\nu}$. In the most quantum region around $x=0$ and in the limit $2GM \gg \ell_{Pl}$ 
we obtain:
\begin{eqnarray}\nonumber
    R &=& \frac{12 \pi}{\Delta} + \Lambda + \mathcal{O} \left[M^{-2/3} \right], \\\nonumber
    R^2 &=& \frac{(12 \pi+\Delta \Lambda)^2}{\Delta^2} + \mathcal{O} \left[M^{-2/3} \right] ,\\\nonumber
    R_{\mu\nu}R^{\mu\nu} &=& \frac{72\pi^2}{\Delta^2} +\frac{4\pi \Lambda}{\Delta}+\frac{\Lambda^2}{2}+ \mathcal{O} \left[M^{-2/3} \right], \\\label{eq:curvat-inv}
    K &=& \frac{144 \pi^2}{\Delta^2}-\frac{8\pi \Lambda}{\Delta}+\Lambda^2 + \mathcal{O} \left[M^{-2/3} \right].
\end{eqnarray}
These expressions are valid for all four spacetimes considered in the previous subsection as long as the approximation $2GM \gg \ell_{Pl}$ holds. We note that for macroscopic black holes, the curvature invariants attain a mass-independent limiting value fully determined by the area gap $\Delta$ and the cosmological constant  $\Lambda$, at the transition surface of the quantum bounce at $x=0$ which replaces the singularity.  Let us recall that the cosmological constant has an upper positive bound given in Eq. \eqref{condition3a} which is of Planck order. Hence, in this case, the curvature invariants at the bounce will never reach super Planckian values. This is not necessarily the case if the cosmological constant is negative. As we see, Ricci scalar can reach unbounded (yet finite) negative values when $(-\Lambda) \gg 12 \pi/\Delta$, while the other scalars above will be positively unbounded (note that they are sums of squares by definition and hence positive even for the case of a large negative cosmological constant).

We now evaluate the curvature invariants in the asymptotic limit, where the classical regime is approached. As expected, their asymptotic expressions in the limit $x \rightarrow \infty$ are given by
\begin{eqnarray}\nonumber
    R &=& 4 \Lambda -\frac{6 (x_0^{min})\Lambda}{x}+ \mathcal{O} \left[\frac{1}{x^2} \right] \label{R_asym}, \\\nonumber 
    R^2 &=& 16 \Lambda^2 -\frac{48 (x_0^{min})\Lambda^2}{x}+ \mathcal{O} \left[\frac{1}{x^2} \right] \label{R_asym}, \\\nonumber 
    R_{\mu\nu}R^{\mu\nu} &=& 4 \Lambda^2 -\frac{12 (x_0^{min})\Lambda^2}{x}+ \mathcal{O} \left[\frac{1}{x^2} \right], \label{R2_asym} \\
    K &=& \frac{8\Lambda^2}{3} -\frac{8 (x_0^{min})\Lambda^2}{x} + \mathcal{O} \left[\frac{1}{x^2} \right]. \label{K_asym}
\end{eqnarray}

\begin{figure}
    \centering
    \includegraphics[width=0.8\linewidth]{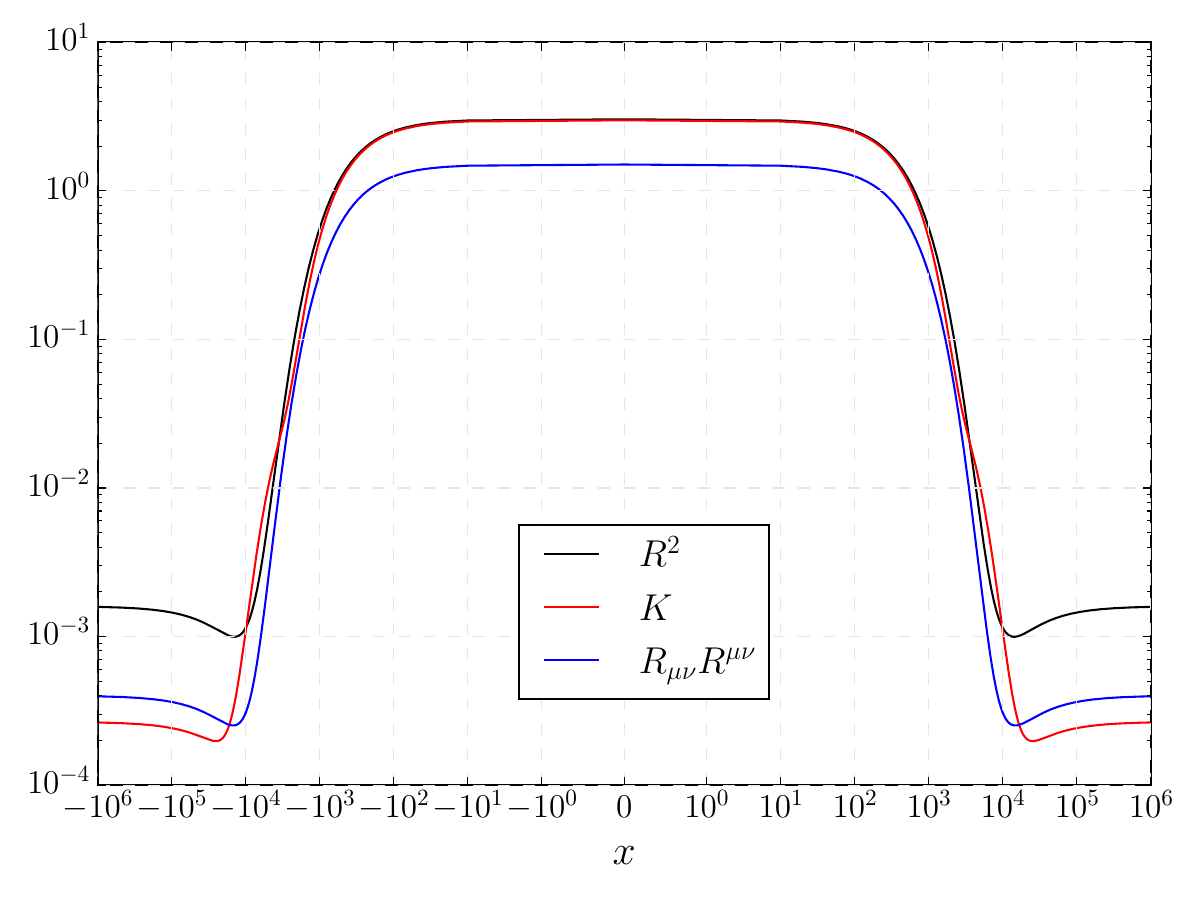}
    \caption{The behaviour of curvature invariants at the transition surface for $M=10^{10}$ and $\Lambda=10^{-2}$.}
    \label{fig:CI}
\end{figure}

These asymptotic expressions are valid for all four possible spacetimes considered in the previous subsection based on the value of $\Lambda$. In particular, we note that the scalar curvature in the asymptotic limit is dominated by $\Lambda$. However, in contrast to the case with $\Lambda=0$ where curvature invariants decay at most as $x^{-6}$ (even if we include charge \cite{BMO2024}) due to quantum corrections, here with $\Lambda\neq 0$, quantum corrections decay only as $x^{-1}$. Hence, we can conclude that a cosmological constant has the unexpected and surprising result of strongly enhancing quantum corrections in the asymptotic region. Note that in the limit $\Lambda \rightarrow 0$, the curvature invariants here reduce to the corresponding expressions in the $\Lambda=0$ case \cite{GOP2020}, and decay at most as $x^{-6}$.

Now, in case of ultramassive spacetimes with $\Lambda>1/9G^2 M^2$ (large cosmological constant), the Planckian bound $\Lambda_{max}$ determined by equation \eqref{condition3a} helps in ensuring that the positive curvature is upper bounded (both at the bounce and) in the asymptotic limit. Thus in case of ultramassive spacetimes, quantum effects may also appear in the asymptotic region if $\Lambda\sim \Lambda_{max}$. For the case of anti-de Sitter spacetime with a central black hole, we have $\Lambda<0$. The curvature in this case is negative and as discussed in the previous section, quantum gravity does not put any bounds on the negative curvature in this case. We are allowed to choose an arbitrarily large (and finite) negative value of the cosmological constant. Still, it must be noted that the curvature remains finite everywhere and the entire spacetime is regular as shown in the Penrose diagram in the Fig. \ref{fig:conf-diagram4}. However, at the transition surface, the curvature invariants will not be bounded by  Planck scale. Instead, their value will be dominated by the negative cosmological constant. This is in contrast with the classical spacetime. There, as one approaches the singularity at $x=0$, curvature blows up and overcomes any contribution from $\Lambda$, no matter how large it is. For illustrative purposes, we show the behaviour of the curvature invariants for a small and positive $\Lambda$ in Fig. \ref{fig:CI}. Since the effective metric is a symmetric function of $x$, the behavior of curvature invariants is also symmetric across the transition surface $x=0$.

\subsection{Effective stress-energy tensor}
It is also enlightening to define an effective stress-energy tensor that captures the effects of quantum corrections. In this point of view, we consider the effective spacetime to be a classical metric governed by Einstein's equations with a cosmological constant $\Lambda$. The effective stress-energy tensor that captures quantum corrections can be easily calculated from the effective metric as 
\begin{eqnarray}
    T_{\mu \nu} = \frac{1}{8\pi G} \left(G_{\mu \nu}+\Lambda {}^{(0)}   g_{\mu \nu}\right),
\end{eqnarray}
where $G_{\mu \nu}$ is the Einstein tensor. From $T_{\mu\nu}$, we can readily extract the effective energy density $\rho$ and the effective radial and tangential pressure densities $p_x$ and $p_{\theta}$ respectively. In the region exterior to the black/white hole horizon, where the effective geometry has a time-like Killing vector (say $t^\mu$), they are given by
\begin{figure}
    \centering
    \includegraphics[width=0.8\linewidth]{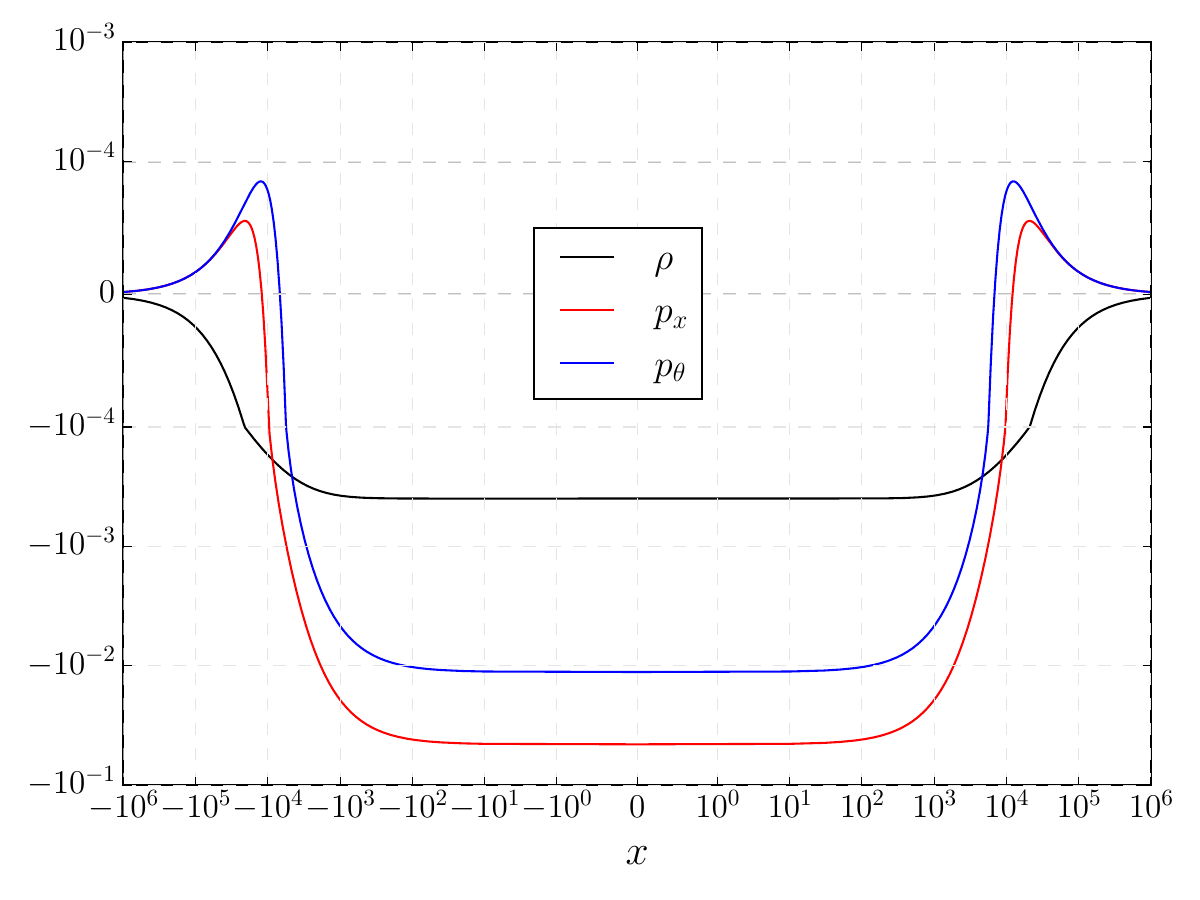}
    \caption{The behaviour of the effective energy density, radial pressure and tangential pressure at the transition surface for $M=10^{10}$ and $\Lambda=10^{-2}$. The scale on the $y$-axis is linear in the range $(-10^{-4},10^{-4})$ and logarithmic in the rest.}
    \label{fig:EP}
\end{figure}
\begin{eqnarray}
    \rho:= \frac{T_{\mu\nu} t^\mu t^\nu}{(-t^\rho t_\rho)}, \\
    p_x := \frac{T_{\mu\nu} r^\mu r^\nu}{ r^\rho r_\rho}, \\
    p_\theta := \frac{T_{\mu\nu} \theta^\mu \theta^\nu}{ \theta^\rho \theta_\rho}.
\end{eqnarray}
where $r^\mu$ is a vector field pointing in the radial direction and $\theta^\mu$ is a vector field pointing in the angular direction. In region bounded by black/white hole horizons, $t^\mu$ becomes space-like while $r^\mu$ becomes time-like, i.e. their roles get reversed.  Since the null energy condition (NEC) takes a simple form in the diagonal gauge, we first change to a diagonal gauge by introducing the transformation $\mathrm{d}\tilde t=\mathrm{d}t+({}^{(0)}g_{tx}/{}^{(0)}g_{tt}) \mathrm{d}x$ in the metric \eqref{effectivemetric}, and choose to evaluate the stress-energy tensor in the diagonal gauge. In the most quantum region at $x=0$, in the limit $2GM \gg \ell_{Pl}$,  they attain a mass-independent limiting value determined by the area gap as follows:
\begin{eqnarray}\nonumber 
    \rho &=& -\frac{\Lambda}{8\pi G} +
    \mathcal{O} \left[M^{-2/3} \right], \\\nonumber 
    p_x &=& -\frac{1}{G \Delta} +\frac{\Lambda}{8\pi G} +  \mathcal{O} \left[M^{-2/3} \right], \\
    p_\theta &=& -\frac{1}{4G \Delta} +\frac{\Lambda}{16\pi G} + \mathcal{O} \left[M^{-2/3} \right].
\end{eqnarray}
On the other hand, their asymptotic behavior at spatial infinity is given by
\begin{eqnarray}\nonumber 
    \rho &=& -\frac{(x_0^{min})\Lambda}{6\pi G x} + \mathcal{O} \left[\frac{1}{x^2} \right], \\\nonumber 
    p_x &=& \frac{(x_0^{min})\Lambda}{4\pi G x} + \mathcal{O} \left[\frac{1}{x^2} \right], \\
    p_\theta &=& \frac{(x_0^{min})\Lambda}{6\pi G x} + \mathcal{O} \left[\frac{1}{x^2} \right],
\end{eqnarray}
showing that quantum corrections to the classical spacetime show a weak fall off compared to the vanishing cosmological constant, which is $x^{-3}$ at most. As we already mentioned, the cosmological constant enhances quantum corrections away from the black hole. We show the behaviour of the energy density and pressure in Fig. \ref{fig:EP}. As discussed in previous subsection, the behavior of energy density and pressures is symmetric across the transition surface $x=0$, due to the symmetric nature of the effective metric.

Finally, note that in this subsection we are viewing the dynamics to be classical while the quantum corrections manifest as an effective stress-energy tensor. Therefore, for the singularity to be resolved, the null energy condition (NEC) must be violated as per the singularity theorems of Penrose \& Hawking. In spherically symmetric spacetimes in the diagonal gauge, the NEC amounts to $\rho+p_x\geq 0$ and $\rho+p_\theta\geq 0$ \cite{nec}. In our effective spacetime, we get at the transition surface and in the limit $2GM \gg \ell_{Pl}$
\begin{equation}
    \rho(x=0)+p_x(x=0) = - \frac{1}{G \Delta}+
    \mathcal{O} \left[M^{-2/3} \right],\quad \rho(x=0)+p_\theta(x=0) = -\frac{1}{4 G \Delta}-\frac{\Lambda }{16 \pi G }+
    \mathcal{O} \left[M^{-2/3} \right]. 
\end{equation}

Hence, the NEC is violated at the transition surface, indicating an avoidance of the singularity in accordance with the singularity theorems. We note that similar plots for NEC violation are obtained near the transition surface for all four distinct spacetimes considered above regardless of the value of $\Lambda$. As per the viewpoint of this subsection, the dynamics of the effective geometry is classical and singularity is resolved due to the violation of the NEC by the effective stress-energy tensor. Alternatively, from the point of view of loop quantum gravity, while the spacetime is vacuum, the quantum corrections are introduced in the geometry itself, which leads to the resolution of the singularity. 

Finally, in the asymptotic limit, we get 
\begin{equation}
    \lim_{x\to \infty}\rho(x)+p_x(x) = +\frac{(x_0^{min})\Lambda}{12\pi G x} +\mathcal{O} \left[\frac{1}{x^2} \right],\quad \rho(x=0)+p_\theta(x=0) = + \frac{(x_0^{min})}{8\pi G x^3} +\mathcal{O} \left[\frac{1}{x^4} \right]. 
\end{equation}
Hence, quantum corrections, despite being non negligible in the asymptotic limit, do not violate the null energy condition. Neither the strong energy condition, although they violate the week and dominant energy conditions since $\rho(x)$ is negative in the asymptotic region.

\subsection{Comparison with other models}

We also want to take the opportunity to draw a comparison of our results with those already reported in the literature about black holes with cosmological constant in LQG. For instance, Ref. \cite{abv} discuses effective geometries of charged black holes with a cosmological constant motivated by a covariant holonomization of the Hamiltonian theory. Their findings agree qualitatively with ours in the sense that these geometries are singularity free, and depending on the relative values of mass, charge and cosmological constant, the global structure of the spacetimes can be at least divided in a similar way: Schwarzschild-de Sitter, extremal and ultramassive spacetimes. Nevertheless, the parameter that controls quantum corrections is left free, and hence curvature scalars in the most quantum region are not expected to reach universal (mass and charge independent) upper bounds. Ref. \cite{linZ} is another work published independently around the same time we submitted ours. It adopts a similar improved dynamics scheme. Their effective geometries share some properties with the ones in our manuscript, but there are some important differences. Concretely, the choice of polymerization of shift function adopted in Ref. \cite{linZ} is different from ours (see Eq. (40) in their paper). This polymerization is of the type $\sin(\rho K_\varphi)\cos(\rho K_\varphi)$ while ours is just of the type $\sin(\rho K_\varphi)$. Our choice is motivated by the polymerization suggested in Ref. \cite{GOP2021}, where it was realized that a choice for the shift like the one in Ref. \cite{linZ}, already considered in Ref. \cite{GOP2020}, turned out to be problematic from the perspective of the covariance of the effective description (see Footnote 1 in Ref. \cite{GOP2020}). Other than that, curvature scalars reach similar universal upper bounds in several curvature scalars, however, their asymptotic behavior is quite different, resulting in a faster decay of quantum corrections.

Finally, we want to mention some recent results about the improved dynamics schemes in effective descriptions of Kantowski-Sachs scenarios, which in the classical theory are diffeomorphic to the black/white hole interior. Concretely, we refer to Refs. \cite{nohoriz1,nohoriz2}. They study the improved dynamics scheme proposed by B\"ohmer-Vandersloot \cite{BV-impr}. Here, the polymer parameters are fixed with two (local) improved dynamics conditions in a diagonal gauge (with a coordinate singularity at the horizon). The conclusions of Refs. \cite{nohoriz1,nohoriz2} are that there are no horizons in the effective spacetime of the B\"ohmer-Vandersloot quantization, being replaced by an infinite tower of transition surfaces. This is in contrast to our results as our quantization does not suffer from this problem. We think that the origin of these discrepancies lies in the B\"ohmer-Vandersloot quantization which considers (local) improved dynamics conditions on a diagonal gauge. The B\"ohmer-Vandersloot quantization works well for the Kantowski-Sachs cosmology \cite{SainiSinghKS}, however, when the vacuum Kantowski-Sachs spacetime is viewed as the interior of a Schwarzschild black hole, there is a coordinate singularity at the horizon in the diagonal gauge where one of the triad variables vanishes. If the improved dynamics conditions are applied locally in the diagonal gauge, they resolve the coordinate singularity at the horizon as if it were a real singularity, as noted earlier in \cite{Saini_thesis} and shown explicitly in \cite{nohoriz1,nohoriz2}. This happens because the
%local application of the improved dynamics scheme in the diagonal gauge ignores} that
triad operators in the loop representation have a discrete spectrum with no vanishing eigenvalues.\footnote{In the physical sector there is superselection on lattices. These lattices typically depend on the choice of the quantum Hamiltonian constraint, that usually excludes from the physical sector the zero eigenvalues of the triads. See e.g. Refs. \cite{aps,mmo}.} Hence, the local application of the improved dynamics conditions in the diagonal gauge is unsuitable in this sense in the context of the Schwarzschild black hole interior. However, it is still unclear how this issue can be satisfactorily solved. Nevertheless, since there is no universal rule for the improved dynamics schemes, it is still possible to explore new proposals aligned with the underlying quantum theory and still providing physically sensible results. For instance, Refs. \cite{aos1,aos2} propose a nonlocal improved dynamics scheme for these geometries where several curvature scalars show mass-independent upper bounds in the most quantum region and without any issue at the horizons, being possible to extend the effective description to the exterior region. Recently, this improved dynamics scheme including a cosmological constant has been studied in Ref. \cite{aos-lambda}. They restrict their study to the interior of the black holes (where curvature is largest) and focus their analysis on the transition from black to white hole geometries (trapped and anti-trapped regions) and their symmetric behavior. However, Ref. \cite{osw} states that these nonlocal improved dynamics schemes might trigger large quantum corrections at small spacetime curvatures. Concretely, the effective description disagrees with the classical one there, and a new black hole horizon is formed. All these models based their conclusions on a well motivated but still postulated effective description.

In our proposal, however, we do have a full quantum theory (of a symmetry reduced model) that allows us to derive effective geometries. They are free from those surprising results about having large quantum corrections in regions where they are not expected. It is also true that we are adopting an improved dynamics scheme where $\delta x$ is obtained by evaluating the improved dynamics condition in the vertex where we expect quantum corrections to be largest (this type of improved dynamics condition aligns with the one in Refs. \cite{aos1,aos2}). Namely, this is not a local improved dynamics condition, as the one studied in Refs. \cite{nohoriz1,nohoriz2}, where one would allow $\delta x$ to take local values on each vertex. Nevertheless, we are presently studying a fully local improved dynamics scheme, and our preliminary results show no issues at the horizon, since $\delta x$ is actually bounded below by $\ell_{Pl}^2/(2 x)$, hence geometrical quantities divided by $\delta x$ will never blow up there. We expect to summarize our definitive results in a future work. 

\section{Conclusions}

In this manuscript, we have extended the spherically symmetric midisuperspace framework of loop quantum gravity using improved dynamics scheme to spherically symmetric black holes with a non-vanishing cosmological constant $\Lambda$, where $\Lambda$ is allowed to range over both positive and negative values. This provides an interesting study of loop quantized black holes in spacetimes that are not asymptotically flat. Our quantization leads to a consistency condition which puts a Planckian upper bound on the possible values of a positive $\Lambda$ which confirms the intuition gained form the analysis of homogeneous spacetimes with a positive $\Lambda$ where a similar bound was obtained. Using semiclassical physical states, we obtained the effective metric which incorporates the leading contributions from quantum corrections. The effective metric is found to be manifestly regular, the singularity having been replaced by a transition surface, beyond which the spacetime can be extended. We show the causal structure of the spacetimes in four distinct cases depending on the value of $\Lambda$, including the case where the spacetime is asymptotically  anti-de Sitter. Concretely, we start with the most physically relevant scenario where the cosmological constant is positive and small, and the mass of the black/white hole is very large compared with Planck scale but sufficiently small compared with the scale provided by the cosmological constant. Here, we find spacetimes containing untrapped regions enclosed by black/white hole and cosmological horizons, with the interior of the black/white hole being regular. The second case analyzed here corresponds to ultramassive spacetimes, where the mass of the black/white hole exceeds the scale provided by the cosmological constant. These spacetimes show no untrapped or null regions, i.e. there are no horizons, and belong to the class of anisotropic cosmological spacetimes. The third case  accounts for a sufficiently large value of the mass such that the black/white horizon meets the cosmological horizon. Here, the untrapped regions collapse to null surfaces. The final case analyzed in the paper corresponds to negative cosmological constant. Here, there is only one horizon associated with the black/white hole, with a regular interior free of singularities. Besides, the quantum theory does not bound the possible values of the negative cosmological constant. All the effective spacetimes we have studied possess various desirable features - (i) quantum effects modify the spacetime only in regions of high curvature inside the black hole horizon, (ii) the Planckian upper bound on $\Lambda>0$ ensures that the curvature in asymptotic region does not exceed Planckian values, (iii) for macroscopic black holes, the curvature invariants approach a mass-independent upper bound at the transition surface which replaces the singularity, (iv) the spacetime is regular at the transition surface which connects the trapped region inside the black hole horizon to an anti-trapped region on the other side, and (v) analysis of effective stress-energy tensor shows violation of null energy condition in the vicinity of the transition surface. 

As a final comment, we want to emphasize a remarkable property we have found while studying the asymptotic limit of the effective geometries of this model. For very large radii, quantum corrections do not decay as rapidly as in the case with a vanishing cosmological constant. This behavior could have profound implications since these corrections, despite being suppressed by Planck scale near the black hole horizon, could be non-negligible at large distances. A complete picture of the properties of these spacetimes will be obtained by a detailed study of the phenomenological aspects in the future, including the case of regular black holes with negative cosmological constant from the perspective of the AdS/CFT conjecture.

\acknowledgments

Financial support is provided by the Spanish Government through the projects PID2020-118159GB-C43, PID2020-119632GB-I00, and PID2022-140831NB-I00 and funded by MI-CIU/AEI/10.13039/50110001103. SS would like to thank Parampreet Singh for helpful discussions. EM was supported by PEDECIBA (Universidad de la República, Uruguay).

\appendix

\section{Quantization: Dynamics}\label{app:qtheory}

 The Abelianization of the Hamiltonian constraint allows us to complete the Dirac quantization in our study. The steps we follow to find the solutions to the Hamiltonian constraint closely follow those of \cite{BMO2024}. We will now promote the Hamiltonian constraint to a quantum operator acting on the kinematical Hilbert space described in section \ref{SecIII}. Starting from the following expression for the Hamiltonian constraint:
\begin{equation}
    H(\underline N) = \int  dx \underline N \left( 2E^\varphi \sqrt{\sqrt{|E^x|}(1+ K^2_{\varphi})- 2GM - E^x \sqrt{|E^x|} \frac{\Lambda}{3} } - (E^x)'(|E^x|)^{1/4} \right).
\end{equation}
we first absorb a factor of $|E^x|^{1/4}$ in the definition of the lapse so we are left with the simpler expression
\begin{equation}
    H(\underline N) = \int  dx \underline N \left( 2E^\varphi \sqrt{(1+ K^2_{\varphi})- \frac{2GM}{\sqrt{|E^x|}} - |E^x| \frac{\Lambda}{3} } - (E^x)' \right).
\end{equation}
We now promote this quantity to a quantum operator by choosing an appropriate ordering:
\begin{align}
\label{total_hamiltonian}
\hat{H}(\underline N) &= \int dx \underline N \left( 2\left[\sqrt{\left( 1 + \widehat{\frac{\sin^2 (\overline{\rho}_j K_\varphi(x_j))  }{\overline{\rho}_j^2}} \right) - \frac{2G\hat M}{\sqrt{\hat{E}^x}} - \hat{E}^x \frac{\Lambda}{3} }\right]\hat{E}^{\varphi} - \left( \hat{E}^x \right)'\right) .
\end{align}
where we have performed the following substitution 
\begin{align}
K_\varphi \rightarrow \frac{\sin (\overline{\rho}_j K_\varphi(x_j))  }{\overline{\rho}_j}
\end{align}
in order to have a well defined operator on the kinematical Hilbert space. We should in principle look for states $\left| \psi \right\rangle$ which are linear combinations of the spin network states $\left| \Vec{k} , \Vec{\mu} , M \right\rangle$ and are annihilated by the Hamiltonian constraint, that is, $\hat{H}(\underline N) \left| \psi \right\rangle = 0$. We begin by noting that this operator acts only on the vertices of the spin network which can be easily seen from the action of the operator $\hat{E}^\varphi$ given in \eqref{Ephi} and from the fact that the action of $\left( \hat{E}^x \right)^\prime$ on a spin network state $\left| \Vec{k} , \Vec{\mu} , M \right\rangle$ is proportional to the difference of the eigenvalues $\ell_P^2 k_j$ corresponding to two different points along the edge, therefore having a non zero contribution only at the vertices of the graph. With this in mind, we can write the Hamiltonian as a sum of operators acting on each vertex of the spin network.
\begin{align}
\label{H_vj}
    \hat{H}(\underline N) = \sum\limits_{v_j} \hat{H}(v_j)
\end{align}
We will now perform a change of representation first. Instead of using the holonomy representation for $K_\varphi$, it is more convenient to use the connection representation; the reason for this is that in the holonomy representation, the operator $\sin (\overline{\rho}_j K_\varphi(x_j))$ introduces a shift in the eigenvalues $\mu$, resulting in a finite difference equation that is not possible to solve in closed form. Instead, if we adopt the connection representation for $K_\varphi$, the operator $\sin (\widehat{\overline{\rho}_j K_\varphi(x_j)})$ acts multiplicatively, while the action of $\hat{E}^\varphi$ is simply $-i \ell_P \partial / \partial K_\varphi$. We are now ready to find the solutions to the Hamiltonian constraint, that is, states of the form
\begin{align}
\label{kinematical_states}
    \left| \Psi \right\rangle = \int \mathrm{d}M \prod\limits_{v_j} \int_0^{\pi/{\Bar{\rho}}_j} \mathrm{d}K_\varphi (v_j)
    \times\sum_{\Vec{k}} \psi(M,\Vec{k},\Vec{K}_\varphi) \left| M,\Vec{k},\Vec{K}_\varphi \right\rangle.
\end{align} 
which are annihilated by \eqref{H_vj}. Given that the Hamiltonian can be written as a sum of operators which act on different vertices, we may write
\begin{align}
\label{kinematical_coefficients}
 \psi(M,Q,\Vec{k},\Vec{K}_\varphi) = \prod\limits_{j} \psi_j (M,k_j,k_{j-1},K_\varphi (v_j)).
\end{align}
Considering now a state of the form \eqref{kinematical_states}, with the coefficients $\psi(M,\Vec{k},\Vec{K}_\varphi)$ given by \eqref{kinematical_coefficients}, and recalling the action of $\hat{E}^x$ given by \eqref{Ex}, acting with the Hamiltonian constraint on such a state yields
\begin{align}
\label{equation_physical_states}
    4 i \ell_P^2 \frac{\sqrt{1+m_j^2 \sin^2 (y_j)}}{m_j} \partial_{y_j} \psi_j + \ell_P^2 (k_j-k_{j-1}) \psi_j = 0,
\end{align}
where we have defined
\begin{align}
\nonumber
    y_j &= \bar{\rho}_j K_\varphi(v_j),
    \\
\nonumber
m_j^2 &= \bar{\rho}_j \left( 1 - \frac{2 G M}{\sqrt{\ell_P^2 k_j}} - \ell_P^2 k_j \frac{\Lambda}{3} \right),
\end{align}
Equation \eqref{equation_physical_states} can be readily solved for $\psi_j$:
\begin{align}
    \psi_j (M,k_j,k_{j-1},K_\varphi (v_j)) =
    \exp{\left(\frac{i}{4} m_j (k_j - k_{j-1}) F(\Bar{\rho}_j K_\varphi (v_j) , i m_j )\right)}
\end{align}
where $F$ is a two variable function defined by
\begin{align}
    F(A,B)=\int_0^A \frac{\mathrm{d}t}{\sqrt{1+B^2 \sin^2 (t)}}
\end{align}
Physical states will then be given by
\begin{align}
\label{physical_states}
| \chi \rangle_{phys} = \int \mathrm{d}M \left| M\right\rangle
\bigotimes_j\left(\sum_{k_j} \chi(k_j) \psi_j (M,k_j,K_{\varphi,j})\left|k_j\right\rangle\right),
\end{align} 
Where the $\chi(k_j)$ are arbitrary functions of norm one on the kinematical Hilbert space. It can be verified that regardless of whether the $m_j$ are real or imaginary, the $\psi_j$ are either pure phases or bounded numbers \cite{GOP2014}, so that the solutions of the Hamiltonian constraint are well defined everywhere. Diffeomorphism invariance can be achieved by applying the usual group averaging procedure. 

Then, a complete set of observables is given by the mass $\hat M$ and the sequence of eigenvalues $k_j$. The later are typically written in a compact form as $\hat O_z$ with eigenvalues $\ell_{\rm Pl}^2 k_{{\rm Int}(Sz)}$, with $z$ being a parameter $z\in[-1,1]$ such that the label $j$ of any edge/vertex is given by $j(z) = {\rm Int}(Sz)$, with $S$ the total number of vertices in the spin network. Despite $z$ is a continuous variable, its relation with $j$ makes the collection of observables $\hat O_z$ being finite, in agreement with $k_j$. 

Therefore, in the kinematical Hilbert space there is a basis of states given by 

\begin{equation}
 \langle \Vec{k}, M |\Vec{k}', M'\rangle=\delta_{\Vec{k}\Vec{k}'}\delta(M-M').
\end{equation}

\end{document}